\documentclass{aa}
\usepackage{graphicx}
\usepackage{amssymb}

\newcommand{\etal}{et al.}

\newcommand{\eg}{e.g.}
\newcommand{\ie}{i.e.}

\newcommand{\Np}{{\cal N}_p}
\newcommand{\Nd}{{\cal N}_d}
\newcommand{\Cl}{C_\ell}
\newcommand{\Cb}{{\cal C}_b}
\newcommand{\alm}{a_{\ell m}}
\newcommand{\Ylm}{Y_{\ell m}}

\newcommand{\DT}{{\Delta T}}
\newcommand{\Dl}{{\Delta \ell}}
\newcommand{\fsky}{f_{\rm sky}}
\newcommand{\rms}{{\it rms}}

\newcommand{\Mll}{M_{\ell\ell^\prime}}
\def\fun#1#2{\lower3.6pt\vbox{\baselineskip0pt\lineskip.9pt
    \ialign{$\mathsurround=0pt#1\hfil##\hfil$\crcr#2\crcr\sim\crcr}}}
\def\lta{\mathrel{\mathpalette\fun <}}    
\def\gta{\mathrel{\mathpalette\fun >}}    

\begin{document}
\title{CMB Power Spectrum Estimation for the Planck Surveyor}

\author{A.~Balbi \inst{1,2} \and G.~de~Gasperis\inst{1} \and
  P.~Natoli\inst{1,2} \and N.~Vittorio\inst{1,2} }

\institute{Dipartimento di Fisica, Universit\`a di Roma ``Tor
  Vergata'', via della Ricerca Scientifica 1, I-00133, Roma, Italy
  \and INFN, Sezione di Roma II, via della Ricerca Scientifica 1,
  I-00133, Roma, Italy }

\authorrunning{A.~Balbi \etal}

\offprints{\email{balbi@roma2.infn.it}}

\date{Received / Accepted}

\abstract{ We use an iterative generalized least squares map-making
  algorithm, in conjunction with Monte Carlo techniques, to obtain
  estimates of the angular power spectrum from cosmic microwave
  background (CMB) maps. This is achieved by characterizing and
  removing the instrumental noise contribution in multipole space.
  This technique produces unbiased estimates and can be applied to an
  arbitrary experiment. In this paper, we use it on realistic
  simulations of Planck Low Frequency Instrument (LFI) observations,
  showing that it can lead to fast and reliable estimation of the CMB
  angular power spectrum from megapixel maps.   }

\maketitle


\section{Introduction}

State of the art measurements of the cosmic microwave background (CMB)
anisotropy are affected by non negligible and (in the case of one
horned experiments) correlated instrumental noise. In order to
minimize this contaminant one resorts to non trivial statistical
techniques which almost invariably require knowledge of the
correlation structure of underlying detector noise, to be measured
from the data themselves. As a consequence, methods to estimate the
noise correlation properties out of time ordered data (TOD) have been
proposed (Dor\'e \etal\ 2001; Stompor \etal\ 2002; Natoli \etal\ 
2002).

Most cosmological information is encoded in the angular power spectrum
of the CMB anisotropies. However, the size of modern CMB datasets and
the presence of correlated instrumental noise in the observations make
it unfeasible to extract the power spectrum from CMB maps by
performing standard and robust matrix manipulations commonly used to
solve linear systems.  Computationally, obtaining a brute force
maximum likelihood estimation of the power spectrum requires at least
$\Np^3$ operations, where $\Np$ is the number of pixels in the map.
Current CMB maps from balloon-borne experiments have $\Np \sim
10^4-10^5$. Brute force power spectrum estimation from these maps is
already prohibitive, and will become totally unattainable for upcoming
space missions such as NASA's MAP
satellite\footnote{http://map.gsfc.nasa.gov} or ESA's Planck
Surveyor\footnote{http://astro.estec.esa.nl/Planck}, whose maps will
have $\Np \gta 10^6$.

A number of strategies have been proposed in the past to address the
problem of power spectrum estimation in a computationally feasible
way.  The MADCAP package (Borrill 1999) uses a parallel implementation
of a quadratic estimator technique (Bond \etal\ 1998) to
obtain maximum likelihood estimates of the power spectrum.  This
method, however, will certainly be too time consuming for future
satellite data sets. Dor\'e \etal\ (2001) applied the same
approach in a hierarchical fashion on subsets of large data sets,
lowering somewhat the computational time requirements.  Szapudi \etal\ 
(2001) adopted an entirely different strategy, extracting the power
spectrum from the 2-point correlation function of the map. Although
this approach is applicable to a generic data set and might in
principle account for noise correlations, until now it has only been
tested in the case of uniform white noise. Finally, other methods were
based on simplifying assumptions tailored on specific experimental
strategies (\eg, Oh \etal\ (1999) for the MAP satellite;
Wandelt \& Hansen (2001) for the Planck Surveyor).

In this work, we focus on characterizing and removing the instrumental
noise contribution in multipole space, in order to obtain unbiased
estimates of the CMB power spectrum.  This approach is based on using
map-making techniques to project estimates of the time stream noise on
the sky, according to the experimental scanning strategy, and on Monte
Carlo (MC) simulations to determine the statistical properties of the
noise in multipole space.  Rather than being based on a noise model or
an approximation (such as, \eg, uncorrelated noise) the time stream
noise properties are estimated directly from the data.  A similar
strategy was developed independently in the MASTER package by Hivon et
al. (2001) and applied to the analysis of the BOOMERanG data
(Netterfield \etal\ 2002).  In this case, fast projection of the noise
on the sky was achieved through non optimal map-making, and filtering
of the data in time domain was required to reduce the correlated noise
contribution.  As such, MASTER does not attempt to create a useful map
as an end-product. Instead, we use the iterative generalized least
squares (IGLS) map-making algorithm described in Natoli \etal\ (2001)
which is fast enough to allow for the generation of an appropriate
number of simulations in a reasonable time.  This algorithm has
optimal statistical properties (i.e. produces unbiased and minimum
variance estimates of the map) and hence it does not require any
filtering of the time stream.  As an application, we use realistic
simulations of Planck Low Frequency Instrument (LFI) observations to
show that we can successfully recover the CMB power spectrum even from
megapixel maps. This is, to date, the first computationally feasible
and unbiased pixel-based power spectrum estimator for Planck which
uses information on the correct (\ie\/ estimated from the data
themselves) instrumental noise covariance.

This paper is organized as follows. In Sect. 2 we discuss the method
we use. In Sect. 3 we present the results of the application to
Planck/LFI. Finally, in Sect. 4 we draw the main conclusions of this
work.


\section{Method}


\subsection{The Statistics of the CMB}\label{statCMB}

We begin by reviewing some basic notions about the statistics of the
CMB on the sky sphere.  The CMB temperature field as a function of the
direction of observation on the sky can be expanded in spherical
harmonics $\Ylm(\theta, \phi)$ as:
\begin{equation}
\DT(\theta,\phi)=\sum_{\ell,m}\alm\Ylm(\theta,\phi).
\end{equation}
The angular power spectrum of the CMB anisotropy is defined as:
\begin{equation}
\Cl=\langle\vert\alm\vert^2\rangle
\end{equation}
where the operation $\langle\cdot\rangle$ represents the average over
the statistical ensemble. We cannot measure this quantity directly,
since our own sky is only a particular realization of the statistical
ensemble. The best possible unbiased estimator of the power spectrum
can be constructed by replacing the ensemble average with an average
over the $2\ell+1$ independent $\alm$ coefficients available for each
$\ell$:
\begin{equation}
\Cl^S={1\over 2\ell+1}\sum_m \left\vert\alm\right\vert^2.\label{clsky}
\end{equation}
We have used the superscript $S$ to indicate that this is just an
estimator of the underlying $\Cl$ obtained from our particular sky
realization.  If each $a_{lm}$ coefficient is a zero-mean Gaussian
variable, as commonly assumed, the $\Cl^S$ estimates follows a
$\chi^2$ distribution with $2\ell +1$ degrees of freedom (DOF) and
\rms\/ (\eg\/ Knox 1995):
\begin{equation}
\Delta\Cl^S = \Cl\sqrt{{2\over 2\ell+1}}.
\end{equation}

If the CMB fluctuations $\DT$ are not measured over the entire sky,
then it is not possible to use the spherical harmonics $\Ylm$ as a
complete basis to perform the multipole expansion. As a consequence,
the power spectrum estimated using formula (\ref{clsky}) will be
biased, and values corresponding to different $\ell$ will be
correlated.  The effect of incomplete sky coverage on the $\Cl$
statistics has been studied by Wandelt \etal\ (2001),
Oh \etal\ (1999), Mortlock et al. (2002).  If
$W(\theta,\phi)=\sum_{\ell,m}w_{\ell m}\Ylm(\theta,\phi)$ is the
weighting function describing a given sky coverage, and:
\begin{equation}
W_\ell={1\over 2\ell +1}\sum_m\vert w_{\ell m}\vert^2
\end{equation}
then it can be shown (see \eg\/ Hivon \etal\ 2001) that:
\begin{equation}
\langle\Cl^S\rangle=\sum_{\ell '} \Mll \Cl
\end{equation}
where:
\begin{equation}
\Mll={2\ell '+1\over 4\pi}\sum_{\ell ''} (2\ell ''+1) W_{\ell ''} \left(
 \begin{array}{ccc}
 \ell & \ell ' & \ell '' \\
 0    & 0      & 0
 \end{array}
\right)^2
\end{equation}
and $\left(
 \begin{array}{ccc}
 \ell & \ell ' & \ell '' \\
 0    & 0      & 0
 \end{array}
\right)$ is the Wigner $3-j$ symbol.

An approximate way to model the effect of partial sky coverage is to
change the number of DOF of the $\Cl$ distribution from $2\ell +1$ to
$(2\ell +1)\fsky$, where $\fsky$ is the observed fraction of the sky.
The $\Cl$ estimates get biased by a $\fsky$ factor, and the \rms\/
becomes (Scott \etal\ 1994; Hobson \& Magueijo 1996):
\begin{equation}
\Delta\Cl^S= \Cl\sqrt{{2\over (2\ell+1)\fsky}}.\label{cosmic_variance}
\end{equation}
Residual correlations in multipole space can be minimized by
estimating the average power spectrum in bands of appropriate width
$\Dl$.

\begin{figure}
  \resizebox{\hsize}{!}  {\includegraphics{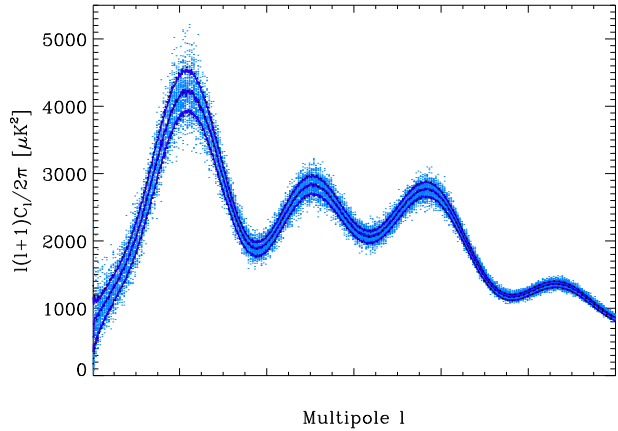}}
\caption{The effect of incomplete sky coverage on the power spectrum estimates. 
  The blue dots are $\Cl/\fsky$ extracted from 100 Monte Carlo
  simulated maps containing only CMB signal, after removing a
  symmetric strip of $\pm 20^\circ$ around the Galactic equator
  ($\fsky=65.8\%$). The black dotted curves are the input theoretical
  $\Cl$ used in the simulations and the upper and lower $1\sigma$
  bounds obtained from Eq. (\ref{cosmic_variance}). The blue
  continuous curves are the mean $\Cl$ and upper and lower $1\sigma$
  bounds estimated from the MC simulations. This shows that Eq.
  (\ref{cosmic_variance}) is a good approximation even when a
  substantial fraction of the sky is unobserved.
}\label{cutsky_effect}
\end{figure}

Since the applications shown in this paper will regard $\Cl$
estimation from nearly full-sky maps ($\fsky\simeq 1$), an exact
treatment of the partial sky coverage effect is totally superfluous,
as the approximation described above yields comparable accuracy.
Indeed, we have verified that this simplified approach provides an
excellent approximation to the exact treatment even when a
considerable fraction of the sky remains unobserved. We have applied a
rather severe Galactic cut ($20^\circ$ symmetric around the equator,
corresponding to $\fsky\approx 65.8\%$) to 100 Monte Carlo simulations
of maps containing only CMB signal, and extracted the corresponding
$\Cl$. Figure \ref{cutsky_effect} shows the results of this test. When
the $\Cl$ estimates are corrected for the $\fsky$ bias they are nearly
undistinguishable from the true underlying theoretical model. From the
same Figure it is also apparent that Eq. (\ref{cosmic_variance})
provides a very good approximation to the true dispersion of the
estimated $\Cl$.

%
%
We then decided, for the present paper, to neglect the complications
of an exact treatment of the partial sky coverage effect. We
emphasize, however, that this does not limit the applicability of the
method, since we are able to treat exactly the case of arbitrary sky
coverage (for example for very small, irregularly shaped patches,
$f_{sky}\ll 1$, as those observed with balloon experiments) by
adopting the correct formalism (as done, \eg, by Hivon \etal\ (2001)).


\subsection{Instrumental Noise Contribution to CMB Power Spectrum Measurements}

The measurements performed by a given experiment can be thought of as
a superposition of sky signal (S) and instrumental noise (N), which
are independent statistical processes. When a map is produced from the
observations, residual noise is left in the pixelized data. This noise
contribution is minimal when a minimum variance map-making (such as
the IGLS) is used to obtain the map.

We can write the observation in pixel $p$ as:
\begin{equation}
m_p = \sum_{\ell,m} \left(\alm^S+\alm^N \right)\Ylm(\theta_p,\phi_p).
\end{equation}
A direct estimation of the power spectrum from such a noisy map would
yield:
\begin{equation}
\Cl^{SN}=
{1\over 2\ell+1}\sum_m\left\vert\left(\alm^S+\alm^N\right)\right\vert^2.
\end{equation}
Expanding this expression, and averaging over the statistical
ensemble, we obtain (due to the independence of the sky and noise
processes):
\begin{equation}
\langle\Cl^{SN}\rangle=\langle\Cl^S\rangle +\langle\Cl^N\rangle.
\end{equation}
Thus, an unbiased estimator of $\Cl^S$ is:
\begin{equation}\label{clest}
\Cl^{E}=\Cl^{SN} - \langle\Cl^N\rangle.
\end{equation}

Direct extraction of the angular power spectrum from a map is a
relatively quick task, that can be performed in $\Np^{3/2}\log \Np$
operations using fast spherical harmonics transforms (Muciaccia \etal\ 1997; 
G{\' o}rski \etal\ 1999).  This
facilitates a fast unbiased estimation of the power spectrum through
Eq. (\ref{clest}), where the noise bias $\langle\Cl^N\rangle$ can be
evaluated using Monte Carlo techniques as discussed in the next
Section.

It can be easily shown that the \rms\/ uncertainty of this power
spectrum estimate is:
\begin{equation}
\Delta\Cl^E=
\left(\Cl+\Cl^N\right)\sqrt{2\over (2\ell +1)\fsky}\label{deltacl}.
\end{equation}

Note that this uncertainty depends on the unknown ensemble average
value $\Cl$. An approximate way of calculating $\Delta\Cl^E$ is simply
to replace $\Cl$ with the estimate $\Cl^E$ in Eq. (\ref{deltacl}). A
more rigorous strategy is to set frequentist confidence intervals by
computing the values of $\Cl$ which are consistent with the estimate
$\Cl^E$ at a given confidence level. We have verified that the two
approaches yield nearly identical 1$\sigma$ error bars for $\Cl^E$.


\section{Results}

To determine the noise properties in multipole space, we first apply
the IGLS map-making procedure on the TOD, ending up with a minimum
variance map of the CMB and an estimate of the true noise properties
in time domain.  We then produce a number of MC realizations of time
streams containing only instrumental noise, where the noise has the
statistical properties measured from the data.  These simulated time
streams naturally incorporate all information about the experiment
observational strategy. We apply again the IGLS map-making algorithm
to these mock data sets, to construct a set of noise maps.  These
noise maps have the same statistical properties of the noise which
contaminates the real map.  They can then be used to characterize the
noise statistical properties in multipole space, for example
estimating the mean $\langle\Cl^N\rangle_{MC}$ to be used in Eqs.\/
(\ref{clest}) and (\ref{deltacl}).

\begin{figure}
  \resizebox{\hsize}{!}  {\includegraphics{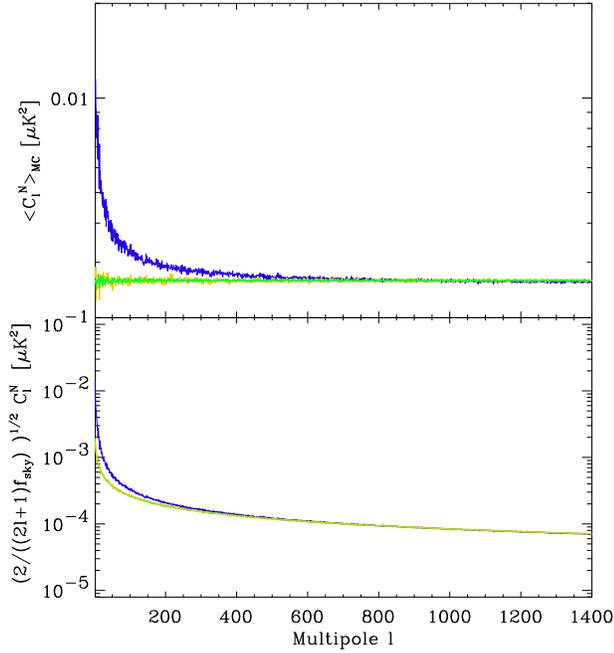}}
\caption{Monte Carlo estimation of the noise properties in multipole 
  space for Planck/LFI. The upper plot shows the average instrumental
  noise contribution to the power spectrum,
  $\langle\Cl^N\rangle_{MC}$. The blue curve was obtained from 22
  realizations of maps containing only instrumental noise with
  realistic properties (including time correlations); the green curve
  is from 100 realizations of non uniform uncorrelated noise; the
  orange curve is from a subset of 22 realizations of the same non
  uniform uncorrelated noise maps. The simulations were performed
  using the nominal sensitivity of all combined 34 receivers at 100
  GHz, for 7 months of observation. The lower plot shows the noise
  contribution to the error bars, from the same sets of simulations.
}\label{noise}
\end{figure}

Figure \ref{noise}\/ shows the result of applying this procedure to
simulated observations of Planck/LFI 100 GHz channel.  The simulations
were produced using the standard Planck scanning strategy (\ie\ a
spinning frequency of 1~r.p.m. and $85^\circ$ offset angle between the
pointing direction and spin axis of the telescope, with the latter
always on the ecliptic) as well as a realistic model of the receiver
noise (\ie\ $1/f$ noise with $f_{\rm knee}=0.1$ Hz and Planck
sensitivity goal, see \eg\ Natoli \etal\ (2002)). A comparison was
made with the case of non uniform uncorrelated noise.  It is apparent
that such a simple model is not accurate enough to characterize the
noise properties in multipole space, particularly for small values of
$\ell$ ($\ell\lta 100$).  It is interesting to observe that, since we
only need to estimate an average from the Monte Carlo, \ie\/ the
$\Cl^N$ to be used in Eqs.\/ (\ref{clest}) and (\ref{deltacl}),
convergence is achieved after a fairly small number of simulations.
This is evident from Fig.  \ref{noise}.
%
%
%

\begin{figure}
  \resizebox{\hsize}{!}  {\includegraphics{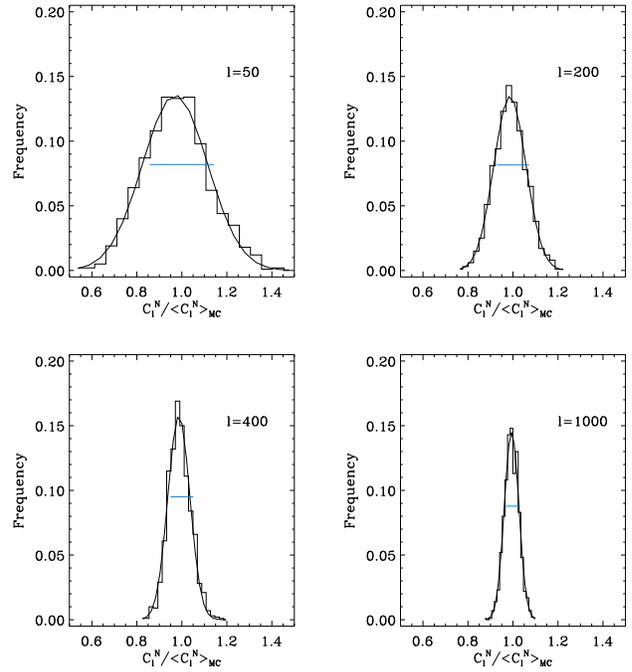}}
\caption{Distribution of the noise contribution in multipole space. 
  The plots show the histogram of $\Cl^N$, normalized to its average
  value, extracted from $1\,000$ Monte Carlo simulations Planck/LFI
  maps, containing only inhomogeneous uncorrelated instrumental noise,
  for four values of $\ell$ (from left to right and from top to
  bottom: $\ell=$50, 200, 400, 1000). Also shown is the $\chi^2$
  distribution with $(2\ell +1)\fsky$ degrees of freedom (continuous
  line) and the corresponding 1$\sigma$ symmetric error bar, given by
  $\sqrt{2/ ((2\ell +1)\fsky)}$ (blue lines).}\label{test_errorbars}
\end{figure}

Figure \ref{test_errorbars}\/ shows histograms of the noise angular
power spectrum $\Cl^N$, for four multipole values, obtained from
$1\,000$ Monte Carlo realizations Planck/LFI maps containing only
instrumental noise. In order to speed up the production of simulated
maps, we used an inhomogeneous white noise model in place of the
correlated noise used in the rest of the paper. This should not alter
the results of this test, since the deviation from the white noise
behaviour is relevant only at low multipoles (see Fig.~\ref{noise}),
where the noise contribution is sub-dominant with respect to the
cosmic variance. The comparison with the theoretical distribution,
\ie\/ a $\chi^2$ with $(2\ell +1)\fsky$ degrees of freedom, shows that
use of Eq.  (\ref{deltacl}) yields nearly optimal error bars for the
power spectrum estimates.
%
%

\begin{figure}[t!]
  \resizebox{\hsize}{!} {\includegraphics{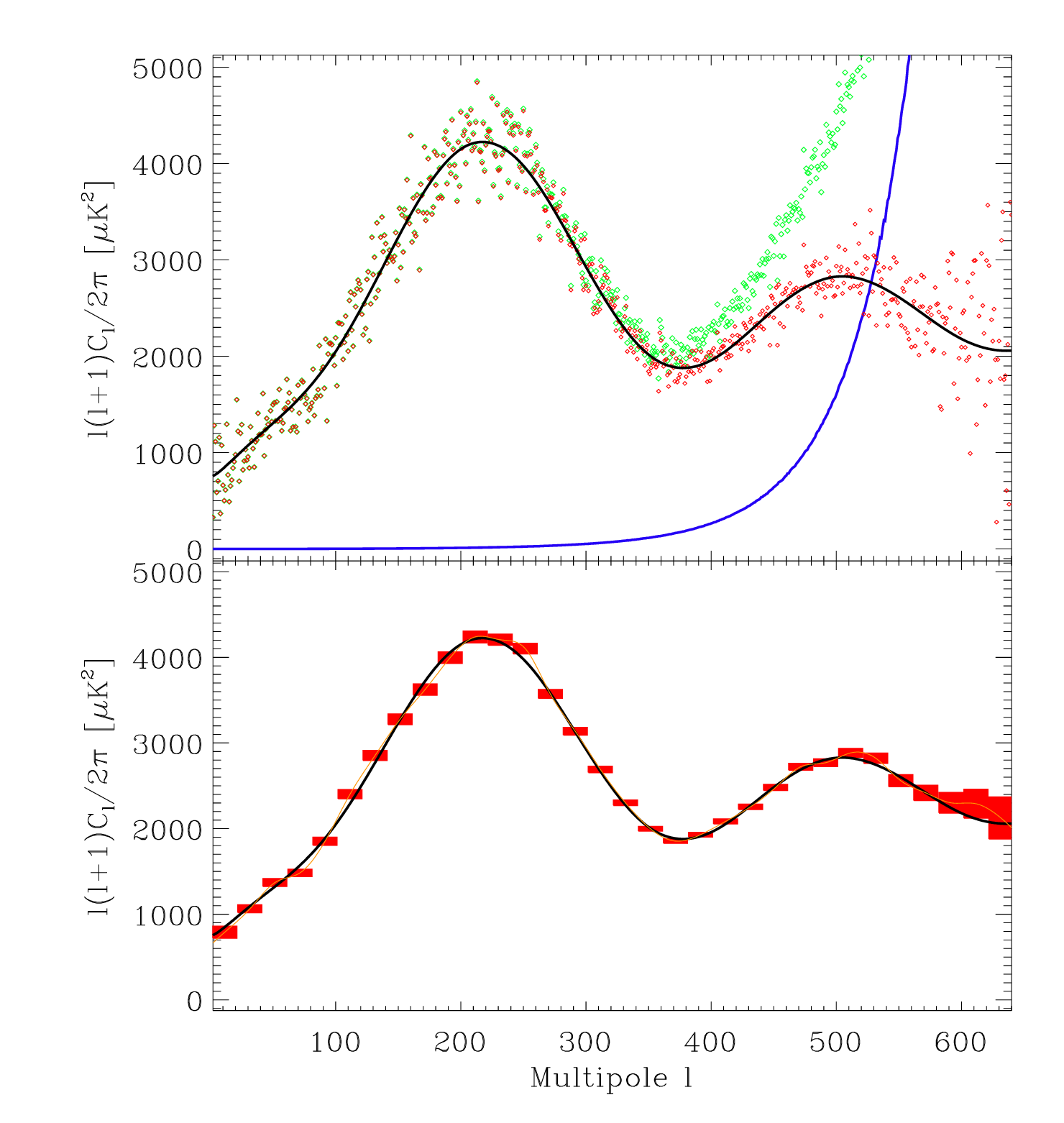}}
\caption{CMB Power spectrum estimation for Planck/LFI 30 GHz channel.
  The simulated data were produced assuming the nominal sensitivity
  from all 4 combined radiometers and 14 months of observation (full
  mission).  The top panel shows the power spectrum estimated directly
  from the map ($\Cl^{SN}$, green points), the MC estimation of the
  noise contribution ($\langle\Cl^{N}\rangle_{MC}$, blue curve), and
  the recovered signal power spectrum ($\Cl^{S}$, red points). The
  bottom panel shows the estimated band-powers $\Cb=\sum_{\ell\in
    b}\ell(\ell+1)\Cl/2\pi$ and their 1$\sigma$ error bars (red
  boxes). The orange curve is the result of a cubic spline
  interpolation of the band-power estimates. In both panels the black
  curve is the input theoretical model used in the simulation (a
  COBE-normalized standard CDM).}\label{Planck_30GHz}
\end{figure}

\begin{figure}[t!]
  \resizebox{\hsize}{!} {\includegraphics{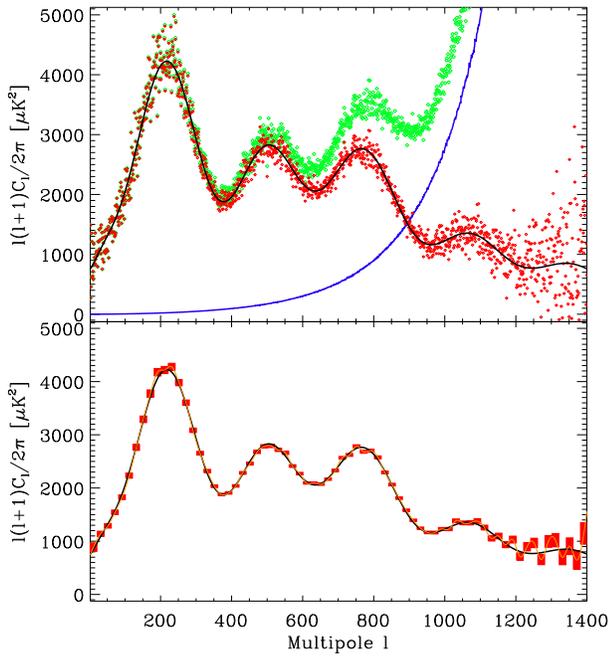}}
\caption{Same as in Fig. \ref{Planck_30GHz}, but for the 34 combined 
  radiometers of Planck/LFI 100 GHz channel and 7 months of
  observation (half mission).}\label{Planck_100GHz}
\end{figure}

Figures \ref{Planck_30GHz} and \ref{Planck_100GHz} show the results of
applying our method to extract the CMB power spectrum from a
simulation of Planck/LFI observations at 30 and 100 GHz. In simulating
the observations we fully took into account the real Planck scanning
strategy. The $85^\circ$ offset between the pointing direction and
spin axis of the telescope leaves unobserved two areas of about 80
square degrees around the ecliptic poles. We modeled the optical
response of the instrument as a symmetric Gaussian beam with FWHM of
33 arcminutes for the 30 GHz detectors and of 10 arcminutes for the
100 GHz detectors. The simulated TOD for the 30 GHz channel has
$\Nd\sim 10^9$ time samples (14 months of observation), which are
mapped into $\Np\sim 7\times 10^5$ pixels. These numbers turn into
$\Nd\sim 2\times 10^9$ (7 months of observation\footnote{Our I/O
  routine is currently able to manage only 32 bit integer. This limits
  the maximum number of time samples we could manage to $2^{31}\simeq
  2\times10^9$. This technical limitation is now being removed.})  and
$\Np\sim 3\times 10^6$ for the 100 GHz channel (we remind that IGLS
map-making scales approximately linearly with $\Nd$).  Going from the
simulated data to the power spectrum estimates took about 12~h for the
30 GHz channel, and about 18~h for the 100 GHz channel, using a
parallel implementation of the IGLS algorithm running on 16 processors
of the Origin 3000 supercomputer at Cineca. These times are dominated
by the production of the MC noise maps, 22 in our case. This number is
enough to produce good estimates for the average noise contribution
(see again Fig. \ref{noise}). For a detailed discussion about memory
requirements, CPU timing, and code scalability of our implementation
of the IGLS map-making algorithm, see the paper by Natoli \etal\/
(2001).

Comparison of our $\Cl$ estimates with the theoretical input model
yields a $\chi^2$/d.o.f. of 1.06 and 1.10 for the 30 and 100 GHz case,
respectively. As mentioned in Section \ref{statCMB}, we also estimated
the power $\Cb=\sum_{\ell\in b}\ell(\ell+1)\Cl/2\pi$ in bands $b$ of
width $\Delta\ell=20$, in order to reduce small spurious correlation
in multipole space. When we use a simple cubic spline algorithm to
interpolate these band-power estimates, the resulting curve is nearly
undistinguishable from the theoretical input model (see bottom panel
in Figs. \ref{Planck_30GHz} and \ref{Planck_100GHz}).
%
%
%
%
%
%
\section{Conclusions}

We have shown that IGLS map-making can be successfully applied, in
conjunction with MC techniques, to the problem of estimating the
angular power spectrum from CMB maps. The method discussed in this
work is fast enough to be already applicable to megapixel maps such as
those expected from the Planck Surveyor. No manipulation of the time
stream (\ie\ high-pass filtering) is required by this method.
Furthermore, no unrealistic simplification of the instrumental noise
behavior is needed, since the noise properties are estimated directly
from the data.  Our estimated noise angular power spectrum contains
the right information on noise correlation, as well as on the details
of scanning strategy, even if we never use explicitly the full pixel
covariance matrix. Although we only presented in this paper an
application to the case of nearly full-sky maps from the Planck
Surveyor, the approach described here can be used for any arbitrary
scanning strategy and sky coverage (for example to analyze balloon
data), as long as the spectral information on the spatial window of
the observation is taken into account.

Finally, we would like to mention a few possible developments of the
work described in this paper. We considered a symmetric beam
approximation in our analysis. This approximation may prove to be too
simplistic when dealing with real experiments. Thus, we are currently
generalizing our map-making algorithm to deal with asymmetric beam
patterns. We also point out that this power spectrum estimation
technique is easily generalizable to CMB polarization maps, since an
implementation of the IGLS map-making algorithm for polarization
observations exists (see Balbi \etal\ 2002).  We shall discuss this
topic in a forthcoming paper.

\begin{acknowledgements} 
  We thank F.K.\ Hansen and D.\ Marinucci for discussions and useful
  advice. The supercomputing resources used for this work have been
  provided by Cineca. We acknowledge use of the HEALPix package
  (http://www.eso.org/science/healpix/).
\end{acknowledgements}



\end{document}